# Discovery of a pair density wave state in a monolayer high-$T_c$ iron-based superconductor

Yanzhao Liu[1,#], Tianheng Wei[1,#], Guanyang He[1], Yi Zhang[2], Ziqiang Wang[3] & Jian Wang[1,4,5,6,*]

[1]*International Center for Quantum Materials, School of Physics, Peking University, Beijing 100871, China*
[2]*Department of Physics, Shanghai University, Shanghai 200444, China*
[3]*Department of Physics, Boston College, Chestnut Hill, MA 0246, USA*
[4]*Collaborative Innovation Center of Quantum Matter, Beijing 100871, China*
[5]*CAS Center for Excellence in Topological Quantum Computation, University of Chinese Academy of Sciences, Beijing 100190, China*
[6]*Beijing Academy of Quantum Information Sciences, Beijing 100193, China*

[#]These authors contribute equally.
*Correspondence to: jianwangphysics@pku.edu.cn (J.W.)

**The pair density wave (PDW) is an extraordinary superconducting state where Cooper pairs carry nonzero momentum (*1,2*). It can emerge when the full condensation of zero momentum Cooper pairs is frustrated. Evidence for the existence of intrinsic PDW order in high-temperature (high-$T_c$) cuprate superconductors (*3,4*) and kagome superconductors (*5*) has emerged recently. However, the PDW order in iron-based high-$T_c$ superconductors has not been observed experimentally. Here, using scanning tunneling microscopy/spectroscopy, we report the discovery of the PDW state in monolayer iron-based high-$T_c$ Fe(Te,Se) films grown on SrTiO$_3$(001) substrates. The PDW state with a period of $\lambda \sim 3.6 a_{Fe}$ ($a_{Fe}$ is the distance between neighboring Fe atoms) is observed at the domain walls by the spatial electronic modulations of the local density of states, superconducting gap, and the $\pi$-phase shift boundaries of the PDW around the dislocations of the intertwined charge density wave order. The discovery of the PDW state in the monolayer Fe(Te,Se) film provides a low-dimensional platform to study the interplay between the correlated electronic states and unconventional Cooper pairing in high-$T_c$ superconductors.**

In a conventional Bardeen-Cooper-Schrieffer (BCS) superconductor, electrons carrying time-reversed momenta form zero center-of-mass momentum Cooper pairs and produce a translation-invariant superconducting order parameter (*6*). In a time-reversal symmetry breaking exchange field, the Cooper pairs may acquire a finite momentum $Q$ and show spatially non-uniform pair density, which is known as the Fulde-Ferrell-Larkin-Ovchinnikov (FFLO) state (*7,8*). Further studies generalized the finite momentum pairing states to the circumstances involving multiple $Q$ components without necessarily breaking the time-reversal symmetry (*9-13*). Such a pair density wave (PDW) state (*1,2,13,14*) is hypothesized to exist in the high-temperature (high-$T_c$) superconducting cuprates and play a fundamental role. Moreover, the orthogonal PDW states in adjacent $CuO_2$ planes are proposed to explain the two-dimensional (2D) superconductivity in bulk $La_{2-x}Ba_xCuO_4$ (x = 1/8) (*10*). However, previous experimental studies of the PDW states were mainly based on bulk superconductors (*3-5,15-17*) and direct visualization of the PDW in the low-dimensional limit is challenging (*18*).

Similar to cuprates, iron-based superconductors possess a complex phase diagram with a variety of symmetry-breaking electronic states, such as the nematic order and stripe antiferromagnetic order (*19-23*). Under in-plane magnetic fields, the anomalous upturn of the upper critical field at low temperatures was reported in iron-based superconductors, which is considered as a signature of the FFLO state (*24,25*). However, direct evidence of the PDW state in iron-based superconductors has been lacking. Among the iron-based superconductors, monolayer FeSe and Fe(Te,Se) films grown on the $SrTiO_3(001)$ (STO) substrates have attracted intensive attention due to the high superconducting transition temperature (*26-32*) and topological electronic structures (*32-35*). Recently, charge ordering has been observed in non-superconducting multilayer FeSe grown on STO (*21,22,36*), which raises a question of whether the PDW can emerge in superconducting monolayer FeSe or Fe(Te,Se) films. Here, using *in situ* scanning tunneling microscopy/spectroscopy (STM/S), we report the observation of the PDW state in one-unit-cell (1-UC) Fe(Te,Se)/STO films along the innate domain walls.

**Domain walls in the 1-UC Fe(Te,Se) film**
The high-quality 1-UC Fe(Te,Se) films were grown on STO substrates by molecular beam epitaxy (MBE) method in an ultrahigh-vacuum system (Methods 'Sample preparation and measurement'). Fig. 1a shows a typical atomically resolved STM topography of the 1-UC Fe(Te,Se) film. The nominal stoichiometry of 1-UC $FeTe_{1-x}Se_x$ (x ~ 0.5) is estimated from the thickness of the second layer film (*28,31,32*) (Extended Data Fig. 1). A typical tunneling spectrum of the 1-UC Fe(Te,Se) film is displayed in Fig. 1b. In this work, all STS measurements were carried out at 4.3 K. The U-shaped feature indicates the fully gapped superconductivity and two pairs of coherence peaks are marked by dashed lines around $\Delta_1 \approx 11$ meV and $\Delta_2 \approx 18$ meV, consistent with previous reports (*31,32*). The spatially uniform superconductivity of the crystalline film is confirmed by the averaged (over ~100 spectra) tunneling spectrum taken along an 11.5 nm line-cut (Extended Data Fig. 1).

A bright line in the topography of the 1-UC Fe(Te,Se) film indicates the domain wall that emerges naturally during the growing process (*37,38*) (Fig. 1c). The orientation of the domain wall is about 45° rotated from the direction of the Te/Se lattice, *i.e.*, along the orientation of the Fe-Fe lattice. Tunneling spectra taken on and off the domain wall show that the superconducting gaps decrease on the domain wall (Fig. 1d and Fig. S1). In some regions of the domain wall, nonzero LDOS can be detected around zero-bias, further confirming the suppression of superconductivity (green curve in Fig. 1d). Comparing with previous STM studies at the domain walls in bulk Fe(Te,Se) (38), we do not observe clear evidence for propagating Majorana states, indicating different domain-wall electronic states in the 1-UC Fe(Te,Se)/STO. Fig. 1e is the zoom-in view of the boxed region shown in Fig. 1c, which characterizes the atomic structure of the domain wall. The grid imprinted in Fig. 1e simulates the Te/Se lattice of the domain on the right hand side. Across the domain wall, a nearly half-unit-cell lattice shift between the grid and the Te/Se lattice of the left domain can be observed (Methods 'Characterization of the domain wall'). The nearly half-unit-cell lattice shift reveals that the Te/Se lattice is locally compressed by about 1/2 $a_{Te/Se}$ across the domain wall ($a_{Te/Se}$ is the distance between neighboring Te/Se atoms), as shown in Fig. 1f.

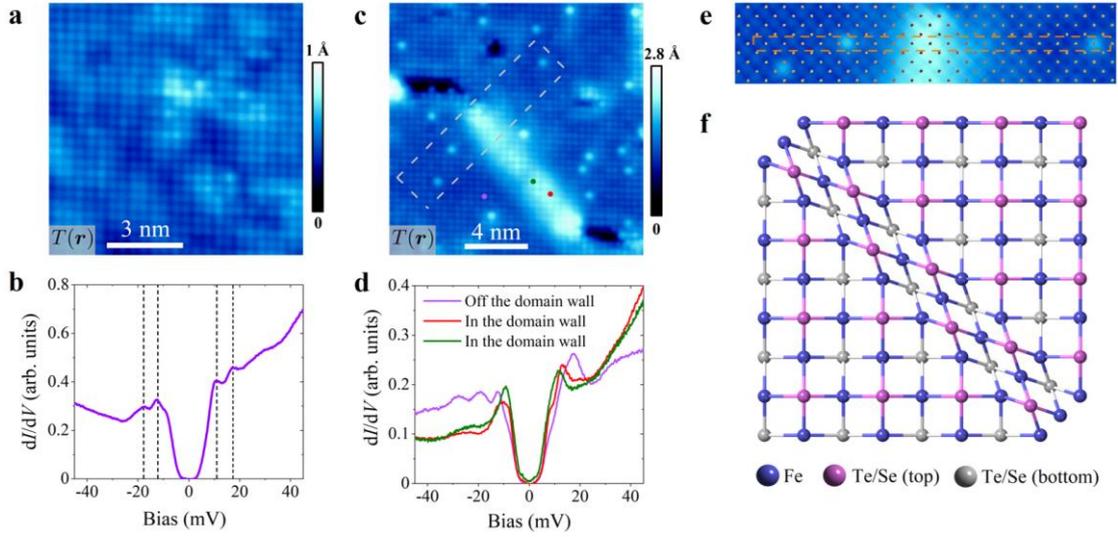

**Fig. 1 | Domain wall along the Fe-Fe bond direction in 1-UC Fe(Te,Se)/STO. a,** An STM topography of the 1-UC Fe(Te,Se)/STO (10×10 nm$^2$, $V_s$ = 0.1 V, $I_s$ = 0.5 nA). **b,** A typical U-shaped tunneling spectrum measured on the 1-UC Fe(Te,Se) film at 4.3 K, which shows a double-gap feature. Superconducting gaps ($\Delta_1 \approx 11$ meV, $\Delta_2 \approx 18$ meV) are indicated by dashed lines. **c,** An STM topographic image of the domain wall structure (bright line) in 1-UC Fe(Te,Se)/STO (16×16 nm$^2$, $V_s$ = 0.1 V, $I_s$ = 0.5 nA ). **d,** Typical tunneling spectra taken on and off the domain wall at 4.3 K. The red, purple, and green curves correspond to the spectra taken at the red, purple, and green points in **c**, respectively. The superconducting gap is smaller on the domain wall, indicating the suppression of the superconductivity. **e,** A zoom-in of the light grey dashed box region in **c**. The spots are illustrative lattices corresponding to the lattice of the right domain. The mismatches between the spots and the lattice of the left domain show a shift around 1/2 $a_{Te/Se}$ across the domain wall. **f,** The schematic of the domain wall in 1-UC Fe(Te,Se)/STO, which illustrates the compression at the domain wall.

**Local density of states modulation at the domain wall**

The electronic properties of the domain wall are examined by tunneling conductance mapping. Fig. 2a shows the drift-corrected and atomically-resolved topography (Methods 'Correction on distortions') of the domain wall (zoom-in view of Fig. 1c) and Fig. 2b displays the corresponding magnitude of the Fourier transform. The zero-bias conductance (ZBC) map $dI/dV(r, V = 0 \text{ mV}) \equiv g(r, V = 0 \text{ mV})$ around the domain wall is shown in Fig. 2c, where $V$ is the bias voltage.

Intriguingly, a spatial modulation of the ZBC $g(r, V = 0 \text{ mV})$ is observed at the domain wall, which is absent in the topography (Figs. 2a-b). This unidirectional LDOS modulation is also absent in regions without the domain wall (Fig. S4). The wavevector of the LDOS modulation, which exists within the superconducting gap, is independent of energy, indicating an origin of electronic order rather than quasiparticle interference (Methods 'Characterization of the LDOS modulation induced by electronic ordering'). In addition to atomic Bragg peaks, the magnitude of the Fourier transform of the ZBC map (Fig. 2d) reveals two Fourier peaks at $Q \sim 0.25 - 0.3 Q_{Fe}$ and $-Q$ ($Q_{Fe}$ is the reciprocal vector of Fe lattice, $Q_{Fe} = 2\pi/a_{Fe}$), corresponding to a unidirectional spatial modulation with a period of $\lambda \sim 3.3 - 4 a_{Fe}$ along the domain wall. The broadening of the Fourier peaks at $Q$ and $Q_{Bragg}$ shown in Fig. 2d results from the spatial confinement of the modulation at the domain wall.

To further characterize the spatial variation of the electronic modulation at wavevector $Q$, a 2D lock-in technique is applied (*4,16,39*) (Methods '2D lock-in technique'). The complex amplitude of *r*-space modulations at $Q$ can be estimated by $A_Q(r) = \int dR A(R) e^{iQ \cdot R} e^{-\frac{(r-R)^2}{2\sigma^2}}$, where $A(R)$ is an arbitrary *r*-space image, and $\sigma$ is the averaging length-scale in *r*-space. The amplitude $|A_Q(r)|$ and phase $\phi_Q^A(r)$ of the modulation at $Q$ can be calculated by the relation $|A_Q(r)| = \sqrt{\text{Re } A_Q(r)^2 + \text{Im } A_Q(r)^2}$ and $\phi_Q^A(r) = \tan^{-1} \frac{\text{Im } A_Q(r)}{\text{Re } A_Q(r)}$, respectively. Figs. 2e-f display the spatial distribution of amplitude $|A_Q(r)|$ and phase $\phi_Q^A(r)$ of the electronic modulation at $Q \sim 0.27 Q_{Fe}$. The value of $Q$ is determined by maximizing the spatial uniformity of the phases at the domain wall (Methods 'Accurate determination of modulation wavevector $Q$'). As shown in Figs. 2e-f, the modulation amplitude at $0.27 Q_{Fe}$ in the ZBC map is confined to the domain wall with uniform phases, consistent with the spatial distribution of the periodic modulation in Fig. 2c.

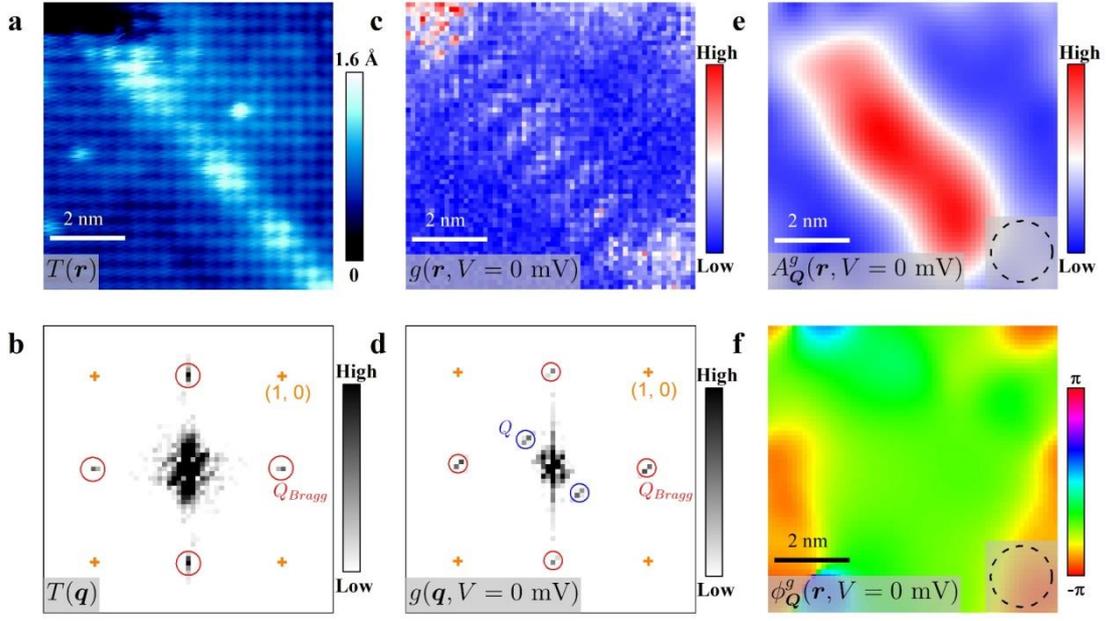

**Fig. 2 | The spatial modulation of the local density of states (LDOS) at the domain wall. a,** Zoom-in image (7.7×7.7 nm$^2$) of Fig. 1c, clearly showing the topography of the domain wall $T(\boldsymbol{r})$. **b,** The magnitude of the Fourier transform $|T(\boldsymbol{q})|$ of **a**. Orange crosses are at $\boldsymbol{q} = (\pm \boldsymbol{Q}_{Fe}, 0), (0, \pm \boldsymbol{Q}_{Fe})$. Bragg peaks of the Te/Se lattice are circled in red. **c,** Zero-bias conductance (ZBC) map $g(\boldsymbol{r}, V = 0 \text{ mV})$ taken at the same area in **a** at 4.3 K, which shows an emergent electronic modulation along the domain wall. **d,** The magnitude of the Fourier transform $g(\boldsymbol{q}, V = 0 \text{ mV})$ of the ZBC map in **c**. The modulation wavevector $\boldsymbol{Q} = 0.25{\sim}0.3\ \boldsymbol{Q}_{Fe}$ circled in blue reveals a spatial modulation with a period around 3.7 $a_{Fe}$ along the direction of the domain wall. **e,** The spatial distribution of the modulation $g_Q(\boldsymbol{r})$ amplitude $|A_Q^g(\boldsymbol{r}, V = 0 \text{ mV})|$ calculated by the two-dimensional lock-in method, which shows that the modulation at $\boldsymbol{Q}$ in $g(\boldsymbol{r}, V = 0 \text{ mV})$ only occurs within the domain wall region. **f,** The spatial distribution of the modulation $g_Q(\boldsymbol{r})$ phase $\phi_Q^g(\boldsymbol{r}, V = 0 \text{ mV})$. The wavevector $\boldsymbol{Q}{\sim}0.27\boldsymbol{Q}_{Fe}$ used in **e** and **f** is determined by analyzing the uniformity of the phase at the domain wall. The averaging length-scales in **e** and **f** are denoted by dashed circles.

**Coherence peak height modulation at the domain wall**

The d$I$/d$V$ maps measured at various bias voltages show that the electronic ordering-induced LDOS modulations exist in the energies within the superconducting gap (Methods 'Characterization of the LDOS modulation induced by electronic ordering'), reminiscent of the PDW state. Previous studies of PDWs in high-$T_c$ cuprates and kagome superconductors have reported a strong correlation between the coherence peak height and the superconducting order parameter (5,40,41). The coherence peak height of the 1-UC Fe(Te,Se) film is measured along the domain wall to provide further evidence for the emergence of the PDW state. Fig. 3a shows the topography of a domain wall. The spatial d$I$/d$V$ map at 2.5 mV and corresponding magnitude of the Fourier transform are shown in Figs. 3b and 3c, respectively. The modulation of LDOS at the domain wall is clearly observed. Further analysis using the 2D lock-in technique reveals a wavevector $\boldsymbol{Q}{\sim}0.28\boldsymbol{Q}_{Fe}$ for the modulation (Figs. 3d-e). Fig. 3f illustrates the tunneling spectra measured along a line-cut parallel to the Fe-Fe bond (light grey arrow in Fig. 3a), which is approximately along the domain wall. The coherence peak height modulation with a period of $\lambda \sim 3.6 a_{Fe}$ can be clearly distinguished (Fig. 3g), consistent with the modulation of LDOS (Figs. 3b-c). The periodic spatial modulation of the coherence peak height is also detected in another domain wall (Fig. S5), further demonstrating the existence of the PDW state. Note that the tunneling spectrum is fully gapped in the 1-UC Fe(Te,Se) film, which is different from other unconventional superconductors exhibiting PDWs, such as cuprates (*3*) and Kagome superconductors (*5*). Although the superconductivity is tempered at the domain wall, the zero-bias conductance $g(r, 0)$ remains much smaller than the coherence peak height. Thus, the gap depth $(g(r, -\Delta) - g(r, 0))$, another characteristic of the PDW, also modulates along the domain wall, as shown in Fig. 3g. Furthermore, the d$I$/d$V$ mapping at -7.5 mV, which is close to the gap edge at the domain wall, is shown in Fig. S6a. The clear periodic pattern and corresponding magnitude of the Fourier transform (Fig. S6b) illustrate the modulation of the coherence peak manifesting the PDW state.

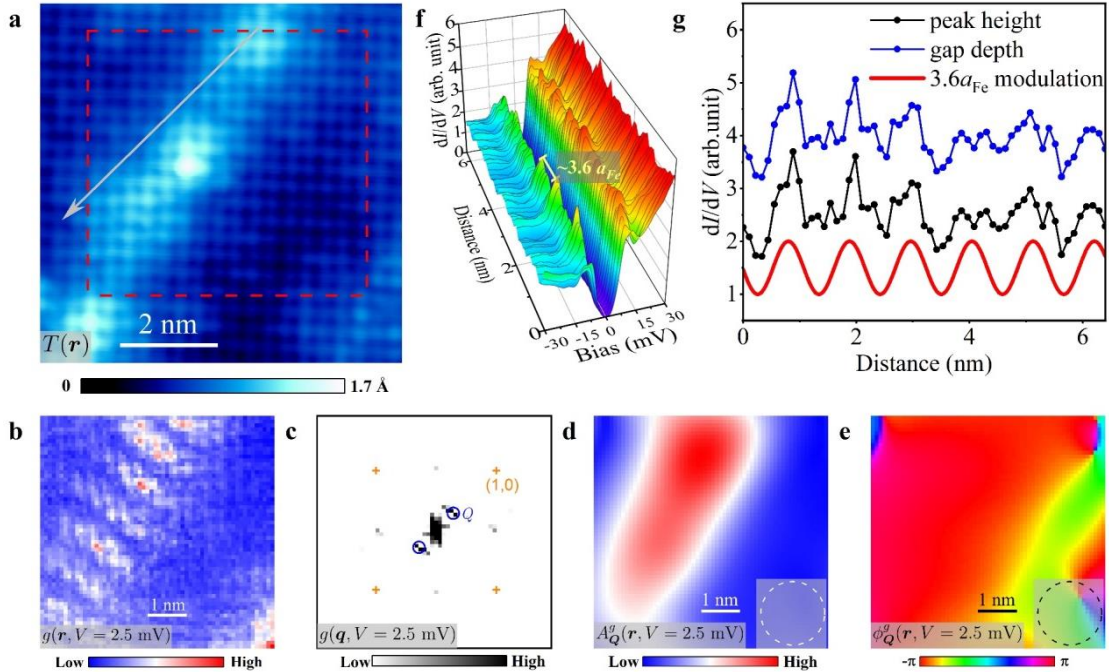

**Fig. 3 | Periodic modulation of the coherence peak height at the domain wall. a,** The STM topography of a domain wall. (8.2×8.2 nm$^2$, $V_s$ = 0.1 V, $I_s$ = 0.5 nA) **b,** d$I$/d$V$ map $g(r, V = 2.5$ mV$)$ taken at the red dashed square in **a** at 4.3 K, in which the LDOS modulation along the domain wall can be observed. **c,** The magnitude of the Fourier transform of **b**. Orange crosses are at $q = (\pm Q_{Fe}, 0), (0, \pm Q_{Fe})$. The modulation wavevector $Q \sim 0.28\, Q_{Fe}$ is marked by solid blue circles. **d, e,** Spatial variation of the amplitude $|A_Q^g(r, V = 2.5$ mV$)|$ and phase $\phi_Q^g(r, V = 2.5$ mV$)$ of the modulation at $Q \sim 0.28\, Q_{Fe}$. The averaging length-scales in **d** and **e** are denoted by dashed circles. **f,** 3D color plot of the d$I$/d$V$ spectra taken along the light grey arrow in **a** ($V_s$ = 0.04 V, $I_s$ = 2.5 nA) at 4.3 K. The cut line is chosen by the Fe-Fe bond direction along the domain wall. The distance in **f** is defined relative to the beginning of the arrow in **a**. The periodic modulation of the coherence peak height is detected, suggesting a pair density wave (PDW) state. **g,** Measured height of coherence peak $g(r, -\Delta)$ and gap depth $(g(r, -\Delta) - g(r, 0))$ in **f**, which is consistent with the simulated modulation (red curve) with a spatial period of $\lambda \sim 3.6\, a_{Fe}$.

**Superconducting gap energy modulation at the domain wall**

Apart from the coherence peak height modulation, the direct observation of superconducting energy gap modulation $\Delta(r)$ has also been recognized as compelling evidence of the multi-$Q$ PDW (*4,5*), in contrast to the helical Fulde-Ferrell (FF) state where only the phase of the order parameter modulates spatially. However, the modulation of $\Delta(r)$ can be challenging to detect in the presence of strong tunneling spectra variations induced by disorder that can obscure the small amplitude of PDW modulations compared to the background signal (*3,4,16*). To investigate the possible gap modulations $\Delta(r)$ and further confirm the PDW state, we first obtain the ZBC $g(r, V = 0 \text{ mV})$ map (Fig. 4a) and the corresponding magnitude of the Fourier transform (Fig. 4b), in the region around a domain wall (Fig. 4a inset). A clear periodic spatial modulation along the domain wall is revealed, consistent with the modulations observed at other domain walls (Figs. 2c, 3b, Extended Data Fig. 2d, Fig. S3e). By extracting the superconducting gap energy from the tunneling conductance spectrum at every pixel in Fig. 4a (Methods 'Measurement of gap modulation'), the spatial distribution of $\Delta_1(r)$, i.e., a gap map is plotted in Fig. 4c, where $\Delta_1$ corresponds to the energy of the smaller superconducting gap. Weak unidirectional modulations are visible and correspond to the two peaks at two $Q{\sim}\pm 0.28Q_{Fe}$ in the Fourier map (Fig. 4d), providing compelling evidence of the PDW state. The broad peak surrounding $Q = 0$ in Fig. 4d is consistent with an approximately uniform background having a wide distribution of the gap values shown in the inset of Fig. 4c. The inset of Fig. 4d plots the line profiles of the Fourier spectra of $g(q, V = 0 \text{ mV})$ and $\Delta_1(q)$ along the (0,0) to (1,0)$Q_{Fe}$ direction (red dashed lines in Fig. 4b and 4d, respectively). Well-defined peaks at $Q{\sim}0.28\,Q_{Fe}$ in both line profiles confirm the spatial modulation of the superconducting gap with the same period as that of LDOS. After the Fourier filtering of Bragg peaks and small-$q$ noise in Fig. 4d, the gap modulation is more evidently depicted in Fig. 4e.

Moreover, the spatial variation of the gap modulation amplitude $|A_Q^{\Delta_1}(r)|$ and phase $\phi_Q^{\Delta_1}(r)$ from the 2D lock-in analysis are displayed in Figs. 4f and 4k, respectively. The amplitude of the gap modulation is confined to the domain wall and accompanied by a uniform phase, further confirming the PDW state at the domain wall in the 1-UC Fe(Te,Se) film.

**Discussion**

The interplay among CDW, PDW, and uniform superconductivity plays a key role in the nature of the PDW phenomenon. The coexistence of a primary CDW order with uniform superconductivity necessarily induces a secondary PDW order. In this case, which has been observed in NbSe$_2$ (*17*), the wavevector $Q$ of the PDW state is the same as that of the CDW. In contrast, in the current 1-UC Fe(Te,Se) films, a primary CDW state is absent (Method 'Characterization of the LDOS modulation induced by electronic ordering'). The observed electronic modulations are thus expected to originate from a primary PDW nucleated at the domain walls. Moreover, a unidirectional PDW state can be either the FF helical state described by the order parameter $\Delta_{PDW}(\mathbf{r}) = \Delta_\mathbf{Q} e^{i\mathbf{Q}\cdot\mathbf{r}}$ with spatially inhomogeneous phase, or the multi-$Q$ state described by $\Delta_{PDW}(\mathbf{r}) = \Delta_\mathbf{Q} e^{i\mathbf{Q}\cdot\mathbf{r}} + \Delta_{-\mathbf{Q}} e^{-i\mathbf{Q}\cdot\mathbf{r}} \propto \cos(\mathbf{Q}\cdot\mathbf{r})$ with spatially periodic modulations of the superconducting gap magnitude (*42*). In the 1-UC Fe(Te,Se) films, the spatial modulation of the SC gap at the domain wall (Figs. 4c and 4d) suggests that the observed PDW state is the multi-$Q$ state, with a wavelength $\lambda{\sim}3.6a_{Fe}$.

To further support this conjecture, we perform more substantiating tests for the existence of a primary multi-$Q$ PDW state. First, a primary multi-$Q$ PDW state is expected to induce a secondary CDW state at wavevector $2Q$ (*11*), as illustrated in Figs. 4g and 4i. We indeed detect a charge density modulation with $2Q$ periodicity, and by analyzing the spatial variation of the amplitude and phase of this secondary $2Q$ CDW, exclude the possible influence of the topmost Te/Se atomic lattice. (Method 'Existence of the secondary CDW state' and Extended Data Figs. 5-6). Second, a π-phase shift in the PDW order around a half-dislocation (black circle in Fig. 4g) is predicted to nucleate at a topological defect with a 2π phase winding (i.e. a vortex) in the induced $2Q$ CDW order (*11*), as indicated by the black circle in Fig. 4i. For a pure PDW state depicted in Fig. 4g, the π-phase shift boundary can be simulated explicitly. Figs. 4h and 4j display the phases of the PDW and the CDW order parameters, showing the π-phase shift boundary in the PDW phase (Fig. 4h) and the corresponding 2π phase-winding around the induced $2Q$ CDW dislocation (Fig. 4j).

Consequently, detecting the π-phase shift boundary in the gap modulations bound to the $2Q$ CDW vortex can provide smoking gun evidence for the emergence of a primary PDW order. Using the 2D lock-in technique, this procedure was attempted in the pioneering work on identifying the PDW in bulk $Bi_2Sr_2CaCu_2O_{8+\delta}$ high-$T_c$ cuprate superconductors (*4*). However, there is a caveat when the PDW order coexists with a uniform superconducting component, as in the bulk of the high-$T_c$ cuprates (*1*). The PDW phase is expected to lock to that of the uniform superconducting order, making the π-phase shift in the pairing gap energetically costly and unlikely to nucleate (*1, 11*). Intriguingly, in the present case, the unidirectional PDW state is only detected at the domain walls in 1-UC Fe(Te,Se)/STO. Recent STM/S studies have reported that a π-phase shift in the superconducting order parameter may exist across the domain boundary in bulk Fe(Te,Se) (*38*). It is thus conceivable that the bulk superconducting order is weakened and prone to inhomogeneous phase shifts at the domain wall boundary in the 1-UC Fe(Te,Se)/STO (Method 'π-phase shift boundaries in the PDW phase'). We are thus motivated to study the π-phase shift in the gap modulation at the domain wall and its spatial correlation to the vortex in the induced $2Q$ CDW.

Applying the 2D lock-in technique to the experimental data, the phase-resolved image of the induced $2Q$ CDW state $\phi_{2Q}^{\rho(2mV)}(r)$ (Method 'Existence of the secondary CDW state') is shown in Fig. 4l, where the topological defects, i.e. double dislocations or vortices, are marked by the black dots. Over the same region, the obtained PDW phase map $\phi_Q^{\Delta_1}(r)$ is shown in Fig. 4k, which is imprinted with the locations of the CDW topological defects (black dots in Fig. 4l). The π-phase shifts in the PDW phase $\phi_Q^{\Delta_1}(r)$, marked by the arrowed cuts in Fig. 4k and inset, are indeed observed near the topological defects of the $2Q$ CDW. More analyses of the π-phase shift boundaries are provided in Method 'π-phase shift boundaries in the PDW phase'. The observation of the π-phase shift boundaries in a PDW bound to dislocations of the induced CDW further confirms the primary multi-$Q$ PDW state in the monolayer Fe-based superconductor.

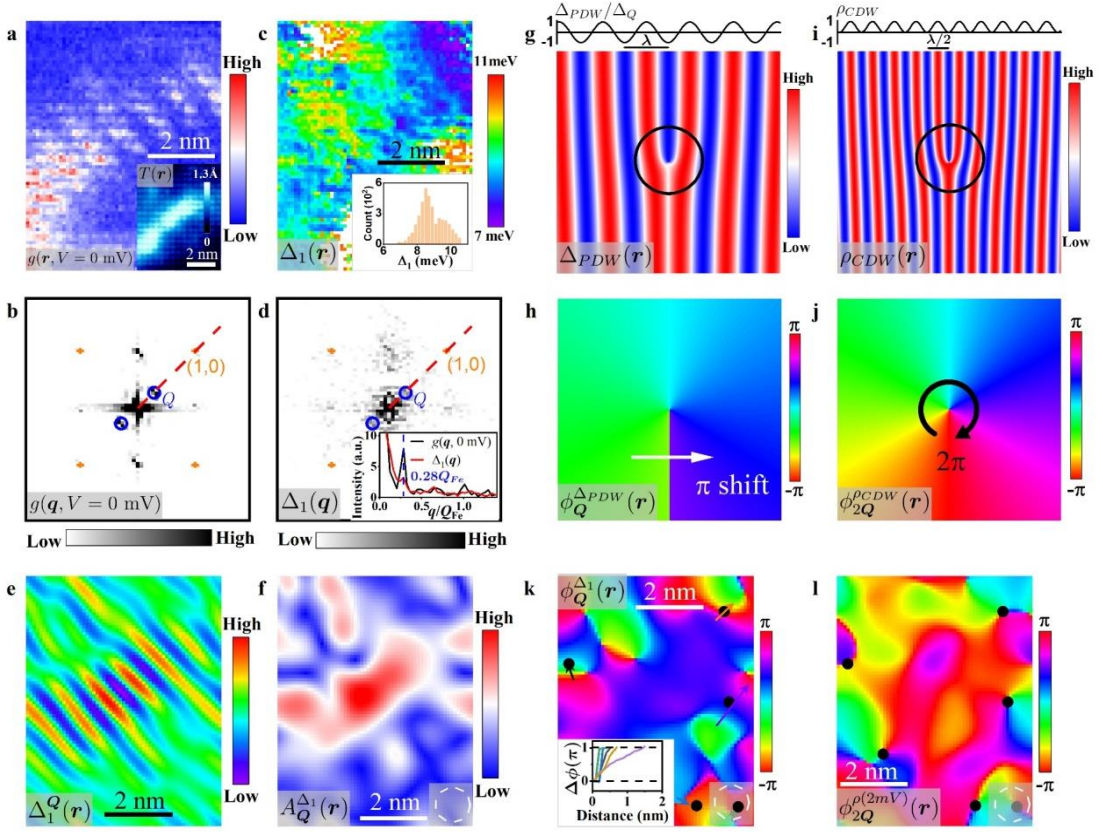

**Fig. 4 | Superconducting gap energy modulations at the domain wall. a,** ZBC map $g(\boldsymbol{r}, V = 0\text{ mV})$ taken around a domain wall (5.9×7.6 nm²) at 4.3 K. The inset shows the topography in the same area ($V_s$ = 0.1 V, $I_s$ = 0.5 nA) **b,** The magnitude of the Fourier transform $g(\boldsymbol{q}, V = 0\text{ mV})$ of **a**. The wavevector circled in blue is at about $\boldsymbol{Q} \sim \pm 0.28\, \boldsymbol{Q}_{Fe}$. **c,** Spatial distribution of $\Delta_1(\boldsymbol{r})$ (gap map) measured in the same area in **a** at 4.3 K. $\Delta_1$ is the energy of the smaller superconducting gap. The inset is the histogram of $\Delta_1$. **d,** The magnitude of the Fourier transform of **c**. Two Fourier peaks at $\boldsymbol{Q} \sim \pm 0.28 \boldsymbol{Q}_{Fe}$ are marked by blue circles, corresponding to the gap modulation along the domain wall. The inset shows the line profiles of $g(\boldsymbol{q}, V = 0\text{ mV})$ and $\Delta_1(\boldsymbol{q})$ along the (0,0) to (1,0)$\boldsymbol{Q}_{Fe}$ direction in **b** and **d**, respectively. Peaks at $\boldsymbol{Q} \sim 0.28\, \boldsymbol{Q}_{Fe}$ appear for both curves. The profiles are normalized for comparison. **e,** The gap map of $\Delta_1(\boldsymbol{r})$ after filtering out Bragg peaks and the small-$\boldsymbol{q}$ noise in **d**, which clearly shows the spatial modulation of the superconducting gap energy at the domain wall. **f,** Spatial variation of the amplitude $|A_{\boldsymbol{Q}}^{\Delta_1}(\boldsymbol{r})|$ of the superconducting gap energy modulation at $\boldsymbol{Q} \sim 0.28\, \boldsymbol{Q}_{Fe}$. The large amplitude at the domain wall justifies the emergence of the modulation at $\boldsymbol{Q}$ in $\Delta_1(\boldsymbol{r})$. **g,** Schematic image of a half-dislocation in the PDW state at $\boldsymbol{Q}$. The half-dislocation is circled in black and the spatial variations of the PDW state along the horizontal axis are plotted at the top. $\Delta_{PDW}$ is the order parameter of the PDW state and $\Delta_Q$ is the amplitude of the PDW state. **h,** Spatial phase of **g**. The π-phase shift boundary is marked by the white arrow. **i,** Schematic image of the CDW state with wavevector 2$\boldsymbol{Q}$ corresponding to the PDW state in **g**. $\rho_{CDW}$ represents the charge density. The topological defect is circled in black and the spatial variations of the CDW state along the horizontal axis are plotted at the top. **j,** Spatial phase of **i**, showing the 2π phase winding of the 2$\boldsymbol{Q}$ CDW vortex. **k,** Spatial variation of the phase $\phi_{\boldsymbol{Q}}^{\Delta_1}(\boldsymbol{r})$ of the superconducting gap energy modulation in **c** at $\boldsymbol{Q} \sim 0.28\, \boldsymbol{Q}_{Fe}$. The inset is the evolution of the phase along the arrows in **k**. **l,** Spatial variation of the phase $\phi_{2\boldsymbol{Q}}^{\rho(2mV)}(\boldsymbol{r})$ of the charge density modulations at 2$\boldsymbol{Q}$.

$\rho(\boldsymbol{r}, 2mV)$ is defined as the difference between tunneling current maps at $\pm 2$ mV. The topological defects which are marked by black dots in **l** are plotted on top of **k**. The averaging length-scales in **f, k, l** are denoted by dashed circles.

What is the origin of the PDW state in 1-UC Fe(Te,Se) films? The appearance of the PDW state confined to the domain walls (Fig. 5a) offers some insights. The local lattice distortion around the domain wall breaks spatial symmetries in the 2D bulk and results in additional strong spin-orbit coupling (SOC). To illustrate how this may generate a quasi-1D PDW along the domain wall, we consider a simple band of the domain wall states, which is split under the SOC by $\boldsymbol{Q}$ in the momentum along the domain wall direction ($k_x$). If time-reversal symmetry is broken, for example by an induced ferromagnetic order around the domain wall, a spin-singlet FF helical state $\Delta_{FF}(\boldsymbol{r}) = \Delta_{\boldsymbol{Q}_z} e^{i\boldsymbol{Q}_z \cdot \boldsymbol{r}}$ can arise from pairing with finite center of mass momentum $\boldsymbol{Q}_z$ on the same Zeeman-shifted Fermi surfaces (Fig. 5b), where $\boldsymbol{Q}_z \approx \frac{E_Z}{\hbar v_F}$ and $E_Z$ is the Zeeman energy and $v_F$ is the Fermi velocity (*43*). However, our findings do not support this scenario for the following reasons. First, the observed PDW wavevector in the 1-UC Fe(Te,Se) films is rather large, $\boldsymbol{Q} \sim 0.28\, \boldsymbol{Q}_{Fe}$. If we choose $\hbar v_F \sim 0.54\, eV \cdot \text{Å}$ (*35*), a large Zeeman energy $E_Z \sim 340$ meV is needed. However, we find no evidence suggesting such a significant spontaneous magnetism around the domain wall. Moreover, the FF state $\Delta_{FF}(\boldsymbol{r})$ favored by the Zeeman scenario only exhibits a single-$\boldsymbol{Q}$ phase modulation and cannot account for the observed superconducting gap modulations.

We therefore propose a different scenario where the time-reversal symmetry is preserved at the domain wall and a multi-$\boldsymbol{Q}$ PDW order created to account for the observed gap modulations. Analogous to the atomic line defect of missing Te/Se atoms, which hosts zero-energy bound states at both ends (*32*), the domain wall can be considered as an embedded quantum structure in the unconventional superconductor where coherent quantum processes and the SOC can generate significant mixed parity pairing (*44*). A single SOC-split band around the zone center, as illustrated in Fig. S7, can indeed produce a finite center of mass momentum, *inter*-FS spin-triplet pairing of the type $\Delta_{\sigma,\boldsymbol{Q}} e^{i\boldsymbol{Q} \cdot \boldsymbol{r}} + \Delta_{-\sigma,-\boldsymbol{Q}} e^{-i\boldsymbol{Q} \cdot \boldsymbol{r}}$, where $\Delta_{\sigma,\boldsymbol{Q}} = \langle c_{k,\sigma} c_{-k+Q,\sigma} \rangle$, $\Delta_{-\sigma,-\boldsymbol{Q}} = \langle c_{k-Q,-\sigma} c_{-k,-\sigma} \rangle$, and $\sigma$ is the spin index. However, such a Kramers FF state (*45*) is still a FF state in that each equal spin pairing channel contains only one momentum, either $+\boldsymbol{Q}$ or $-\boldsymbol{Q}$, and cannot produce the superconducting gap modulations. We must thus consider a minimal band structure for the domain wall states that contains two valleys, which can arise from projecting the electron pockets at X and Y points in the Brillouin zone of the pristine 1-UC Fe(Te,Se) (*35*) onto the domain wall along the Fe-Fe direction, as shown in Figs. 5c and 5d. Note that since the two valleys are separately by momentum $\pi$, the SOC leads to two sets of oppositely spin-split bands depicted in Fig. 5d. Remarkably, this more realistic electronic structure is capable of producing a multi-$\boldsymbol{Q}$ PDW state. The intra-valley, finite momentum equal-spin pairing $\Delta_{\sigma,\pm\boldsymbol{Q}} = \langle c_{k,\sigma} c_{-k\pm Q,\sigma} \rangle$ at $\pm\boldsymbol{Q}$ leads to $\Delta^{\sigma}_{PDW}(\boldsymbol{r}) = \Delta_\sigma \cos \boldsymbol{Q} \cdot \boldsymbol{r}$, which coexists with intra-valley uniform pairing of opposite spins (Fig. S8) and is consistent with our observation of a time-reversal invariant domain wall PDW. Clearly, the microscopic description of the origin and properties of the domain-wall PDW requires further theoretical and experimental investigations.

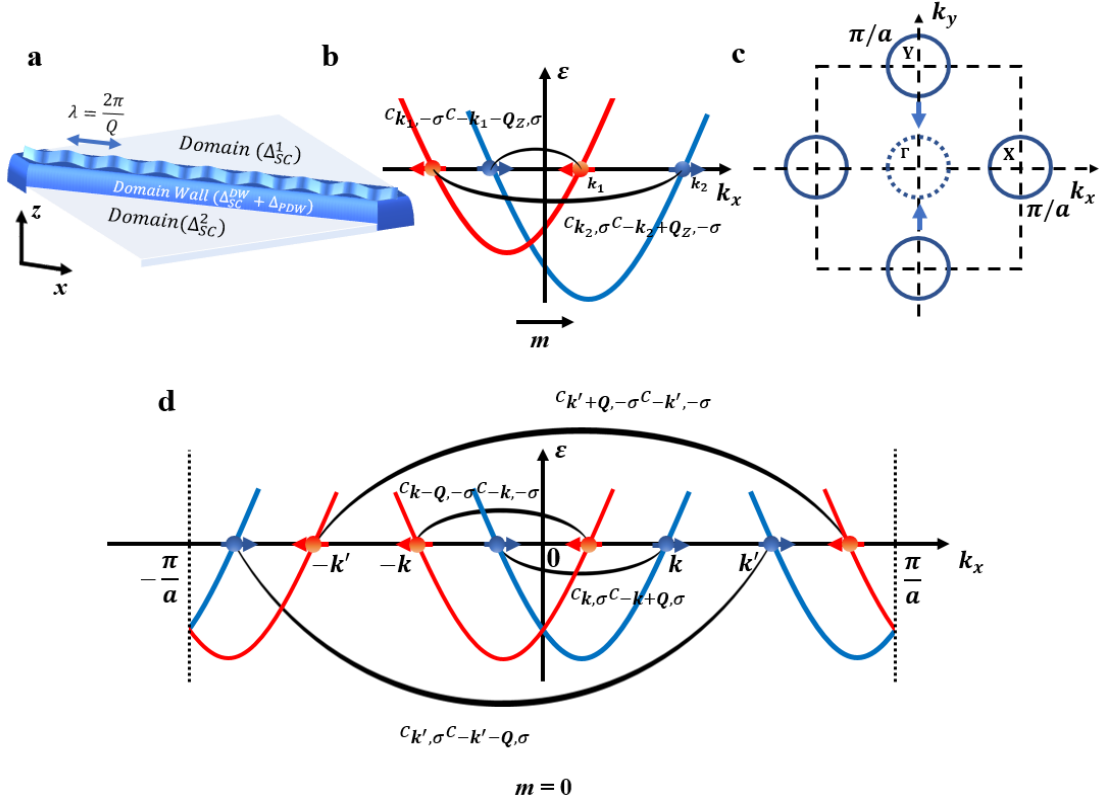

**Fig. 5 | Scenarios for the PDW state at the domain wall. a**, Schematic of the PDW state at the domain wall boundary. $\Delta_{SC}^{1,2}$ and $\Delta_{SC}^{DW}$ represent the zero-momentum superconductivity in the domains (1 and 2) and at the domain wall, respectively. **b,** Illustration of the PDW state induced by the spin-orbit coupling (SOC) coexisting with the Zeeman field ($m$) along $x$ direction. $\langle c_{k_1,-\sigma} c_{-k_1-Q_Z,\sigma} \rangle$ and $\langle c_{k_2,\sigma} c_{-k_2+Q_Z,-\sigma} \rangle$ are the spin-singlet pairing order parameters with the finite momentum $Q_Z$ induced by Zeeman field, and $\sigma$ is the spin index. **c**, Schematic Fermi surface of monolayer Fe(Te,Se) shown in one-Fe unit-cell picture. Four solid circles at X(Y) points represent the electron pockets and the dashed circle at Γ point is the projected pocket for 1D domain wall. **d,** Illustration of the PDW state induced by the intra-valley, *inter*-Fermi surface pairing. $\langle c_{k,\sigma} c_{-k+Q,\sigma} \rangle$, $\langle c_{k',\sigma} c_{-k-Q,\sigma} \rangle$, $\langle c_{k-Q,-\sigma} c_{-k,-\sigma} \rangle$ and $\langle c_{k'+Q,-\sigma} c_{-k',-\sigma} \rangle$ are the spin-triplet order parameters with the finite momenta $\pm Q$ induced by the SOC. Red and blue curves in **b** and **d** plot the energy bands ($\varepsilon(k_x)$) with opposite spin orientations indicated by red and blue arrows, respectively.

In summary, we have detected an incommensurate PDW state with a period of $\lambda \sim 3.6 a_{Fe}$ located at the domain walls in the 1-UC Fe(Te,Se)/STO. The primary PDW state is evidenced by the periodic spatial electronic modulations, such as the LDOS, coherence peak height, gap depth and superconducting gap size at the domain wall, as well as the π-phase shift boundary in the PDW phase near the vortex of the induced 2**Q** CDW. These findings are consistent with an intriguing intra-valley, finite-momentum pairing across the SOC split FSs in the odd-parity, spin-triplet pairing channel. Our findings demonstrate that quantum structures such as naturally-appearing domain walls embedded in 2D iron-based superconductors provide a new material platform to study the PDW state and its interplay with the topological electronic states and unconventional high-$T_c$ superconductivity.

**Acknowledgements** We acknowledge technical assistance from Xiaotong Xu and Cheng Chen and discussions with Yanan Li. This work was supported by the National Natural Science Foundation of China (Grant No. 11888101), the National Key Research and Development Program of China (Grant No. 2018YFA0305604), the Beijing Natural Science Foundation (Z180010), the Strategic Priority Research Program of Chinese Academy of Sciences (XDB28000000), the China Postdoctoral Science Foundation (2021M700253). Z.W. is supported by the U.S. Department of Energy, Basic Energy Sciences Grant No. DE-FG02-99ER45747.

**Author contributions** The order of the first two authors was determined arbitrarily. J.W. conceived and supervised the research. Y.L. and T.W. grew the samples and analyzed the experimental data. Y.L., T.W. and G.H. carried out the STM/STS experiments. Z.W. and Y. Z. proposed the theoretical model. Y.L., T.W., Z.W. and J.W. wrote the manuscript.

**Competing interests** The authors declare no competing interests.

**Data availability** Data measured or analyzed during this study are available from the corresponding author on reasonable request.

## Methods

**Sample preparation and measurement**

Our experiments were carried out in an ultrahigh-vacuum MBE-STM system (Scienta Omicron). The Nb-doped SrTiO$_3$(001) (wt 0.7%) substrates were thermally boiled in deionized water at 90°C for 50min and chemically etched in 12% hydrochloric acid for 45 min. Then the substrates were transported into the MBE chamber with a base pressure about $2 \times 10^{-10}$ mbar and underwent Se-flux treatment at 950°C to expose an atomically-flat TiO$_2$-terminated surface. High-quality monolayer FeTe$_{1-x}$Se$_x$ films were grown on pretreated Nb-doped SrTiO$_3$ substrates by co-depositing high-purity Fe(99.994%), Te(99.999%) and Se(99.999%) while the substrates kept at 340°C. Then the films were annealed at 380°C for 3h and transported to the *in situ* STM chamber after annealing. The thickness of the 2$^{nd}$-UC FeTe$_{1-x}$Se$_x$ film is approximately 0.59 nm (Extended Data Fig. 1), corresponding to the composition $x \approx 0.5$ (*28*). All the results shown were acquired at 4.3 K with a polycrystalline PtIr tip by the standard lock-in technique. The modulation voltage is $V_{\text{mod}} = 1$ mV at 1.7595 kHz. The set-up of all results is $V_s = 0.04$ V, $I_s = 2.5$ nA unless stated otherwise.

**Characterization of the domain wall**

We characterized the crystal structure of the domain wall by analyzing the top Te/Se lattice on both sides of the domain wall. The lattice in each domain can be described by $T_i(\boldsymbol{r}) = \cos(\boldsymbol{q}_{i,x} \cdot \boldsymbol{r} + \theta_{i,x}) + \cos(\boldsymbol{q}_{i,y} \cdot \boldsymbol{r} + \theta_{i,y})$, where $i = 1,2$ labels two domains, $\boldsymbol{q}_{x(y)}$ is the reciprocal lattice vector of the top Te/Se lattice along $x(y)$ direction and $\theta$ is the phase of the lattice. In the area far from the domain wall, the reciprocal lattice vectors of top Te/Se lattice are the same, i.e. $\boldsymbol{q}_{1,x(y)} = \boldsymbol{q}_{2,x(y)}$. As shown in Extended Data Fig. 2, regions as large as possible inside each domain but far from the domain wall are selected, in which $\boldsymbol{q}_{1,x(y)} = \boldsymbol{q}_{2,x(y)}$. The periodically perfect lattices calculated using the $\boldsymbol{q}_{x(y)}$ and $\theta_{i,x(y)}$ are shown by white dots in Extended Data Fig. 2. The phase difference $\Delta\theta_{x(y)} = \theta_{1,x(y)} - \theta_{2,x(y)}$ between the two domains represents a displacement along $x(y)$ direction, which turns out to be a compression along Fe-Fe direction at the domain wall. Besides, we also calculated the phase maps using the 2D lock-in technique around the Bragg peaks (*38*), as shown in Fig. S2. The phase shifts obtained by these two methods are consistent.

**Correction on distortions in topography $T(\boldsymbol{r})$ and differential conductance map $g(\boldsymbol{r}, V)$**

The spectroscopic imaging STM experiments are time-consuming, thus the piezo creep (and many other effects) usually causes remarkable distortions on the data set. In the well-known Lawler-Fujita algorithm (*46*), the total displacement field $\boldsymbol{u}(\boldsymbol{r})$ is considered to be roughly a constant over a coarsening length scale $1/\Lambda_u$ such that $\Lambda_u \ll Q_{Bragg}$. However, in some of our data sets, $\boldsymbol{u}(\boldsymbol{r})$ varies quickly with space, leading to a large $\Lambda_u$ that doesn't satisfy the condition $\Lambda_u \ll Q_{Bragg}$. The distortion information is broadened to a comparatively large range in $\boldsymbol{q}$-space and it is hard to define unambiguous Bragg peaks. Besides, the large distortions usually result in the severely wrapped phase in the Lawler-Fujita algorithm. Thus, it is not suitable to directly use the Lawler-Fujita method to correct these data sets. Considering that the distortions caused by piezo creep are accumulated with time, we subtracted a creep-caused displacement assuming that the creep exponentially decays with time. After this pretreatment, we applied the Lawler-Fujita algorithm to eliminate the residual picometer-scale distortions. In this way, we calculated $\boldsymbol{u}(\boldsymbol{r})$ for topography and corrected $T(r)$ and $g(r,V)$ by subtracting $u(r)$, as shown in Fig. S3.

**Characterization of the LDOS modulation induced by electronic ordering**

The electronic ordering origin of the LDOS modulations can be confirmed by the differential conductance map $g(r,V)$ with sweeping bias. The possibility of the LDOS modulations in $g(r,V)$ originating from quasiparticle interference (QPI) can be excluded. Quasiparticles scattered by impurities interference with each other, leading to the spatially modulated LDOS. The wavevectors of the QPI pattern connect the constant energy contours (CECs) of quasiparticles with energy $E = e \cdot V$. The CECs vary with energy $E$, thus the wavevectors of QPI are dispersive with bias $V$ (*37*). As shown in Extended Data Fig. 3b, the LDOS modulations at domain walls in our work are non-dispersive within the superconducting gap, which is decisive evidence of electronic ordering rather than QPI origin.

Recently, the smectic phase has been detected in uniaxial strained LiFeAs (*47*) and the correlation between the charge-ordered stripes and local anisotropic strain has been studied in FeSe multilayers grown on STO (*36*). This raises the question of whether the PDW state in the 1-UC Fe(Te,Se) films is induced by a primary CDW state coexisting with superconductivity. The primary CDW state usually appears as spatial modulations in STM topographic images and the CDW-like stripe order has been observed in the topography of ~ 30-UC FeSe films grown on STO (*21*). However, no spatial modulation can be detected in the topography of our 1-UC Fe(Te,Se) films. Moreover, the stripe order in FeSe multilayers can be detected in the d$I$/d$V$ mapping at a large bias (≥100 mV) (*21,22,36*), which is another characteristic of the primary CDW state. As shown in Extended data Fig. 3, the PDW state observed in our 1-UC Fe(Te,Se) films only emerges at the energy within the superconducting gap (~10 meV), which further excludes the possibility of the primary CDW induced PDW state in 1-UC Fe(Te,Se) films.

**2D lock-in technique**

Consider a real space image containing a series of modulation wavevectors: $A(r) = \sum_Q a_Q(r) e^{-iQ \cdot r}$, where $a_Q(r)$ is the complex amplitude for wavevector $Q$ at position $r$. Then the Fourier transform of $A(r)$ ($F[A(r)]$) can be written as:

$$A(q) \equiv F[A(r)] = \sum_Q \int dq' a_Q(q') \delta(Q - q + q') = \sum_Q a_Q(q - Q)$$

where $a_Q(q - Q)$ is the Fourier transform of the complex amplitude centered at $Q$. $a_Q(q)$ is almost zero for $|q| > \Lambda_Q$, when $a_Q(r)$ is roughly a constant over a coarsening length scale $1/\Lambda_Q$ ($\Lambda_Q \ll \Delta Q$ and $\Delta Q$ represents a typical separation between different wavevectors). In this case, the Fourier transform of different wavevectors' complex amplitudes are clearly separated. $a_Q(q)$ can be approximately extracted by shifting it back to the center and multiplying a Gaussian window with cut-off length $\sigma_q$ ($\Lambda_Q < \sigma_q \ll \Delta Q$), which is the inverse of averaging length-scale in $r$-space. Then approximate complex amplitude in real space $A_Q(r)$ can be obtained by inverse Fourier transform. In practice, the 2D lock-in technique (*4*) is realized as below:

$$A_Q(r) = F^{-1}[A_Q(q)] = F^{-1}[F(A(r)e^{iQ \cdot r}) \cdot \frac{1}{\sqrt{2\pi}\sigma_q} e^{-\frac{q^2}{2\sigma_q^2}}]$$

which is equivalent to the process stated above. Using the 2D lock-in technique (*4*), the PDW

wavevector is carefully analyzed, as shown in the main text.

**Accurate determination of modulation wavevector $Q$**

For a real space image $A(r)$ in a limited field of view (FOV) $L_1 \times L_2$, the discrete Fourier transform $A(q)$ has a pixel size of $\frac{2\pi}{L_1} \times \frac{2\pi}{L_2}$. Limited by the domain wall size, the FOVs in our experiments are typically smaller than 16nm×16nm, which intensely lower the resolution in $q$-space. Since the PDW state is limited at the domain wall, the inhomogeneous distribution of the modulation amplitude also makes the peaks in $q$-space broaden to a small region around the accurate wavevector. Thus, it is hard to directly get the accurate wavevector $Q$ using Fourier transformation. To determine the wavevector $Q$ accurately, we used the 2D lock-in technique (*4*) to solve the phase map under a tentative $Q_t$.

$$A_{Q_t}(r) = F^{-1}[A_{Q_t}(q)] = F^{-1}[F(A(r)e^{iQ_t \cdot r}) \cdot \frac{1}{\sqrt{2\pi}\sigma_q} e^{-\frac{q^2}{2\sigma_q^2}}]$$

$$\phi_{Q_t}^A(r) = \tan^{-1}\left(\frac{\text{Im}(A_{Q_t}(r))}{\text{Re}(A_{Q_t}(r))}\right)$$

where $\sigma_q$ is the $q$-space cut-off length. $\sigma_q$ is carefully chosen to exclude the influence of other wavevectors, which contain the low-frequency information of the slowly varied modulation amplitude and phase. For $Q_t \sim Q$,

$$\phi_{Q_t}^A(r) \approx (Q_t - Q) \cdot r + \phi$$

which is nearly a constant only when $Q_t = Q$. As shown in Extended Data Fig. 4, by counting the density distribution of $\phi_{Q_t}^A(r)$ at the domain wall for a series of tentative $Q_t$, we determine the accurate wavevector $Q$ which corresponds to the sharpest peak in $\phi_{Q_t}^A(r_{\text{at the domain wall}})$ distribution histogram ($r_{\text{at the domain wall}}$ represents the area of domain wall).

**Measurement of gap modulation**

We measured tunneling spectra at each position $r$, and estimate the smaller gap $\Delta_1(r)$ as follows:
1. Take the second derivative of the differential conductance spectra;
2. Find the bias $V_0$ that has the minimal value of $\frac{d^2}{dV^2}g(r,V)$;
3. Use three data points of $\frac{d^2}{dV^2}g(r,V)$ in the neighbor of $V_0$, namely $\{V_{-1}, V_0, V_{+1}\}$, to fit a quadratic function and take the apex $eV_\Delta$ as the gap value $\Delta(r)$ (*16*).
4. Select the smaller gap values obtained at each position as $\Delta_1$. A critical energy gap $\Delta_c \sim 10.5$ meV is used as the upper limit of $\Delta_1$, and $\Delta_c$ is determined by the statistical histogram of gaps (dashed line in Fig. S9).

Then the gap modulation was analyzed by calculating the Fourier transform $\Delta_1(q)$ and 2D lock-in signals $|A_Q^{\Delta_1}(r)|$ and $\phi_Q^{\Delta_1}(r)$.

**Existence of the secondary CDW state**

Considering the case of coexisting zero-momentum superconducting and PDW order parameters,

namely $\Delta(r) = \Delta_{SC} + \Delta_Q e^{iQ \cdot r} + \Delta_{-Q} e^{-iQ \cdot r}$, CDW order is allowed by the interactions between different order parameters. The product of PDW order parameters induces the secondary CDW order

$$\rho_{\pm 2Q}^{CDW} = \rho_{\mp 2Q}^{CDW*} \propto \Delta_{\pm Q} \Delta_{\mp Q}^*$$

To reveal the existence of the $2Q$ CDW order, we calculated the spatial variation of the phase $\phi_{2Q}^{\rho(V)}(r)$ and the amplitude $\left|A_{2Q}^{\rho(V)}(r)\right|$ of the $\rho(r, V)$ modulation at $2Q$. The $\rho(r, V)$ map is defined as the difference between tunneling currents at $\pm V$

$$\rho(r, V) = I(r, V) - I(r, -V) \propto \int_{-V}^{V} N(r, E) dE.$$

As shown in Extended Data Fig. 5, the amplitude of the $2Q$ modulation is gradually restricted to the domain wall region and the phase of the modulation becomes more uniform in the domain wall region with decreasing bias voltage, indicating the emergence of the $2Q$ CDW order.

In Fe(Te,Se) films, the wavevector of the PDW state is about 0.27 $Q_{Fe}$ and the wavevector of the secondary $2Q$ CDW should be close to 0.54 $Q_{Fe}$. However, the topmost Te/Se atomic lattice shows a reciprocal lattice vector of 0.5$Q_{Fe}$ when projected to the Fe-Fe direction, which is close to $2Q$ and may mix with the $2Q$ CDW. As shown in Extended Data Fig. 5, the energy range where the $2Q$ CDW exists is very small (< 8 meV), which cannot result from the lattice distortion existing in a large energy range.

To further distinguish the $2Q$ CDW signal from the Te/Se atomic lattice projection, we calculated the spatial variation of the phase $\phi_{0.5Q_{Fe}}^{T}(r)$ (Extended Data Fig. 6a) and the amplitude $\left|A_{0.5Q_{Fe}}^{T}(r)\right|$ of the STM topography $T(r)$ (Extended Data Fig. 6c) at precisely 0.5 $Q_{Fe}$. The set point of the STM topography is 40 mV, where the contribution of the $2Q$ CDW modulation is negligible.

As shown in the phase map $\phi_{0.5Q_{Fe}}^{T}(r)$ (Extended Data Fig. 6a), the phase of 0.5$Q_{Fe}$ modulation in the topography (i.e., topmost Te/Se atomic lattice) is not distributed uniformly at the domain wall, which is different from the uniform phase of $2Q$ CDW $\phi_{2Q}^{\rho(2mV)}(r)$ (Extended Data Fig. 6b). Furthermore, the amplitude of 0.5$Q_{Fe}$ modulation $\left|A_{0.5Q_{Fe}}^{T}(r)\right|$ are also randomly distributed in the whole field of view (FOV) (Extended Data Fig. 6c), while the amplitude of $2Q$ CDW $\left|A_{2Q}^{\rho(2mV)}(r)\right|$ (Extended Data Fig. 6d) is restricted to the domain wall area, which is consistent with the uniform phase $\phi_{2Q}^{\rho(2mV)}(r)$. The topological defects with $2\pi$ phase winding are further marked in $\phi_{0.5Q_{Fe}}^{T}(r)$ and $\phi_{2Q}^{\rho(2mV)}(r)$ (black dots in Extended Data Figs. 6a-b). The topological defects in $\phi_{0.5Q_{Fe}}^{T}(r)$ are randomly distributed in the whole FOV, while the topological defects in $\phi_{2Q}^{\rho(2mV)}(r)$ are mainly at the boundary of the domain wall. These results indicate that the $2Q$ CDW is independent of the atomic lattice.

**π-phase shift boundaries in the PDW phase**

Consider the PDW order $(\Delta_Q, \Delta_{-Q}) = \Delta_{PDW}(e^{i\varphi_Q}, e^{i\varphi_{-Q}})$ coexists with the uniform superconducting order $\Delta_0 e^{i\theta}$, where $\theta$ and $\varphi_{\pm Q}$ are U(1) phases. The whole phases of $Q$ and $-Q$ PDW order parameters $\Delta_{\pm Q}$ can be expressed as $(Q \cdot r + \varphi_Q(r))$ and $(-Q \cdot r + \varphi_{-Q}(r))$. In STM experiments, the magnitude of the order parameter is detected, which can be expressed as:

$$|\Delta| = |\Delta_0 e^{i\theta} + \Delta_Q e^{iQ \cdot r} + \Delta_{-Q} e^{-iQ \cdot r}|$$
$$= |\Delta_0| + 2|\Delta_{PDW}| \cos\left(\theta - \frac{\varphi_Q + \varphi_{-Q}}{2}\right) \cos\left(Q \cdot r + \frac{\varphi_Q - \varphi_{-Q}}{2}\right).$$

One-half of the relative phase between $Q$ and $-Q$ PDW order parameters ($\phi_{STM}(r) = Q \cdot r + \frac{\varphi_Q(r) - \varphi_{-Q}(r)}{2}$) is extracted from the second term, which spatially modulates as $\cos(\phi_{STM}(r))$. During the 2D lock-in calculation process, a relative phase $Q \cdot (r - r_0)$ is subtracted and the phase value in the obtained phase map is $\phi(r) = \frac{\varphi_Q(r) - \varphi_{-Q}(r)}{2} + Q \cdot r_0$, which depends on the selected origin ($r_0$). Note that the phase shift between two points $\phi(r_1) - \phi(r_2)$ or along a closed path $\oint d\phi$ is independent of $r_0$. Thus, the phase shift, which contains the information of topological features, is meaningful and discussed in our work ($Q \cdot r_0$ can be ignored).

We further introduce $\gamma = (\varphi_Q + \varphi_{-Q})/2$ that appears in the supercurrent formula, and $Q \cdot d \equiv \phi(r) = (\varphi_Q - \varphi_{-Q})/2$, which is caused by the dislocation ($d$). For pure PDW case ($\Delta_0 = 0$), the π-phase winding of $Q \cdot d$ is observable by STM and corresponds to 2π-phase winding in $2Q$ CDW order. The continuity of order parameters requires $\varphi_{\pm Q} = \gamma \pm Q \cdot d$ must wind by integer multiples of 2π, which implies π-phase shift of $\gamma$ and leads to what was often referred to as a half-vortex.

As shown in Figs. 4k and 4l, the π-phase shift boundaries in PDW phase ($Q \cdot d$) are located near the topological defects with 2π-phase winding in $2Q$ CDW phase, which is reminiscent of the half-vortices in pure PDW phase (*11*). But, in the presence of a uniform superconducting component, the lowest coupling term between the PDW and superconducting order parameters is given by:

$$f_c = \beta_{c1} \Delta_0^2 \Delta_{PDW}^2 + \beta_{c2} \Delta_0^2 \Delta_{PDW}^2 \cos(2\theta - 2\gamma),$$

where $\beta_{c1}$ and $\beta_{c2}$ are expansion coefficients. To minimize the free energy, the second term locks $\gamma = \theta$ (or $\theta + \pi$) and $\gamma = \theta \pm \pi/2$ for $\beta_{c2} < 0$ and $\beta_{c2} > 0$, respectively. When $\theta$ is uniform for the superconducting order, the phase $\gamma$ (($\varphi_Q + \varphi_{-Q})/2$) should also be uniform, which prohibits the emergence of the half-vortices with π-phase winding. Meanwhile, the π-phase shift in $Q \cdot d$ observed in the STM experiments is also be hindered due to the requirement of continuous order parameters.

The existence of π-phase shift in $Q \cdot d$ of the 1-UC Fe(Te,Se)/STO may be explained by the non-uniform superconducting phase at the domain walls. Similar to the superconducting phase at the domain boundary of bulk Fe(Te,Se) (*38*), the superconducting order may not be uniform at the domain wall boundary in the 1-UC Fe(Te,Se)/STO. The π-phase shift of the PDW phase can be allowed if a π-phase shift of superconducting order appears at the domain wall boundary, which leads to topological defects with 2π phase winding in $2Q$ CDW phase.

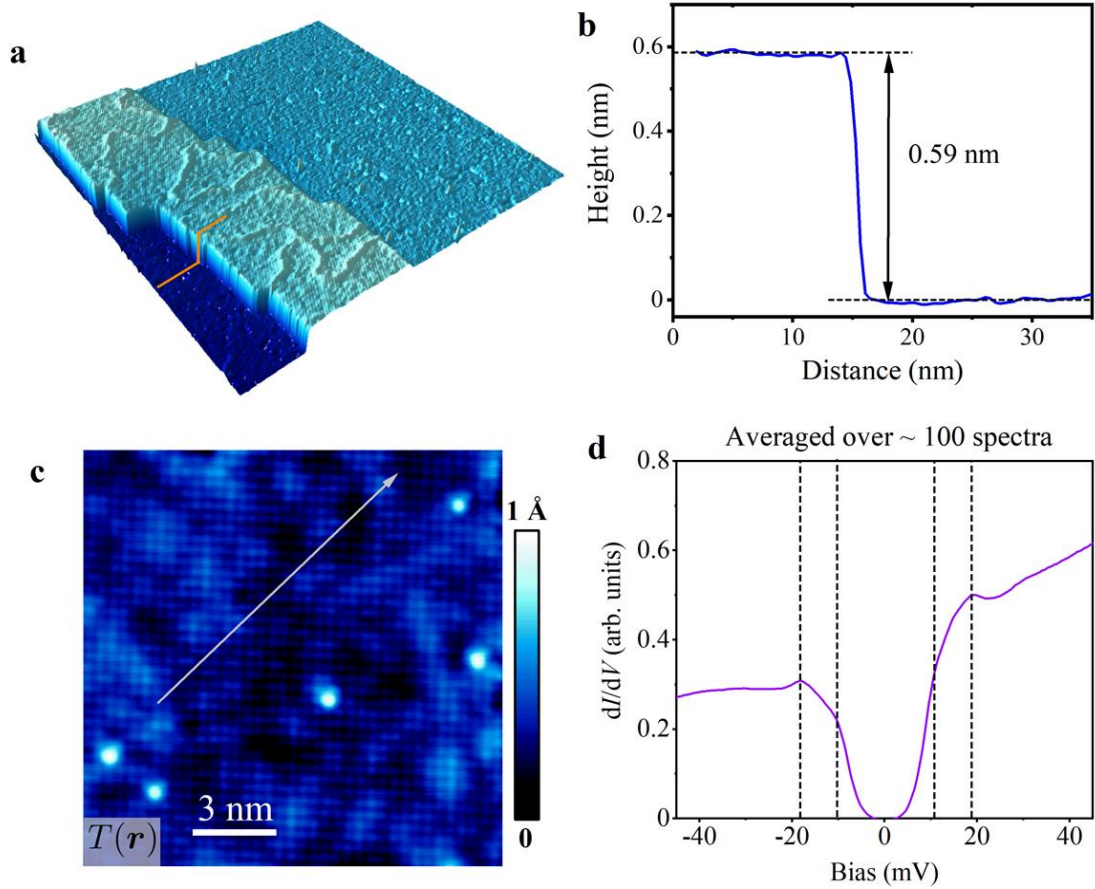

**Extended Data Fig. 1 | More information about the 1-UC Fe(Te,Se)/STO. a,** An STM topography of 1-2 UC Fe(Te,Se)/STO at large-scale (200×200 nm$^2$, $V_s$ = 1 V, $I_s$ = 0.5 nA) with terraces of the STO substrate (lighter color means the higher height). The cyan area at the edge of the lower STO terrace shows the 2$^{nd}$-UC Fe(Te,Se) film. **b,** Line profile taken along the orange curve in **a**. The thickness of the 1-UC FeTe$_{1-x}$Se$_x$ film is approximately 0.59 nm, corresponding to the composition $x \approx 0.5$. **c,** An STM image of 1-UC Fe(Te,Se)/STO (10×10 nm$^2$, $V_s$ = 0.1 V, $I_s$ = 0.5 nA). **d,** Averaged tunneling spectrum (averaged over ~100 spectra) taken along the light grey arrow in **c**, which shows two superconducting gaps of $\Delta_1 \approx 11$ meV and $\Delta_2 \approx 19$ meV.

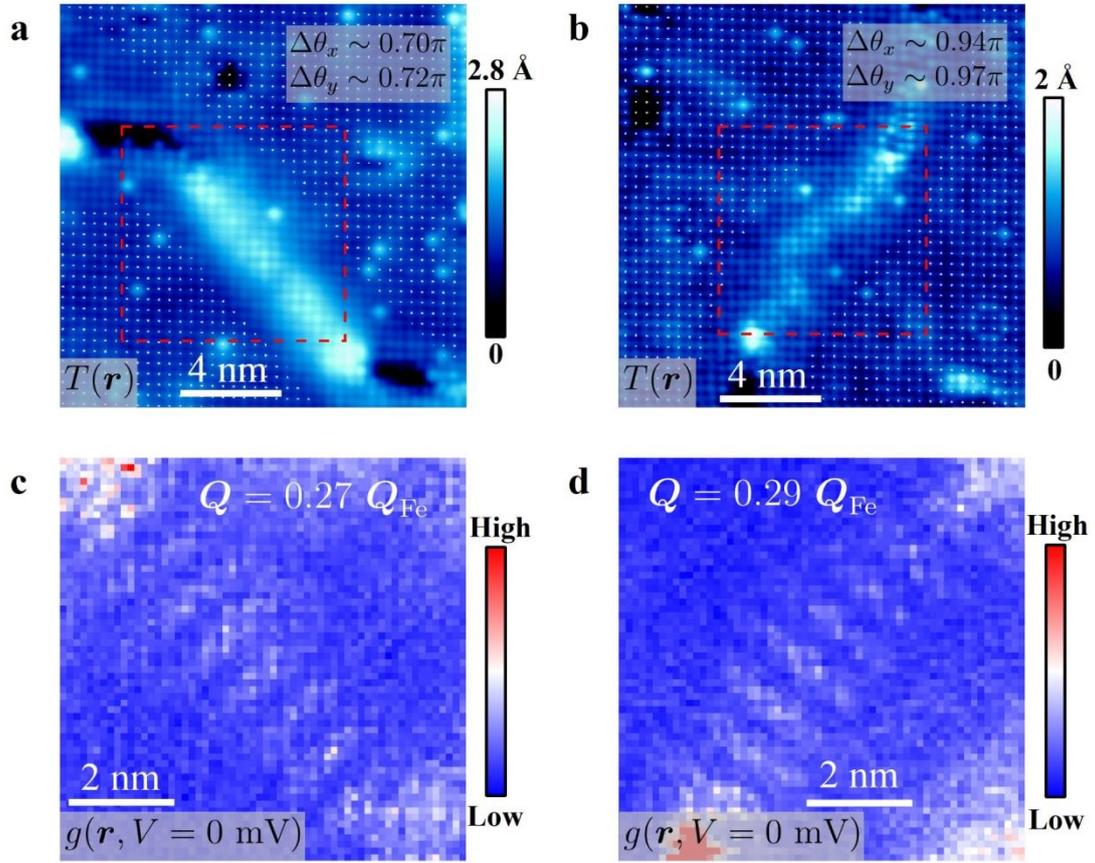

**Extended Data Fig. 2 | Characterization of domain walls. a, b,** STM images (16×16 nm$^2$, $V_s$ = 0.1 V, $I_s$ = 0.5 nA) of two domain walls. **c, d,** d$I$/d$V$ maps of two domain walls. The lattice vectors $q_{i,x(y)}$ of two regions marked by white spots in **a** and **b** satisfy $q_{1,x(y)} = q_{2,x(y)}$. d$I$/d$V$ maps shown in **c** and **d** are measured at the red dashed box in **a** and **b**, respectively.

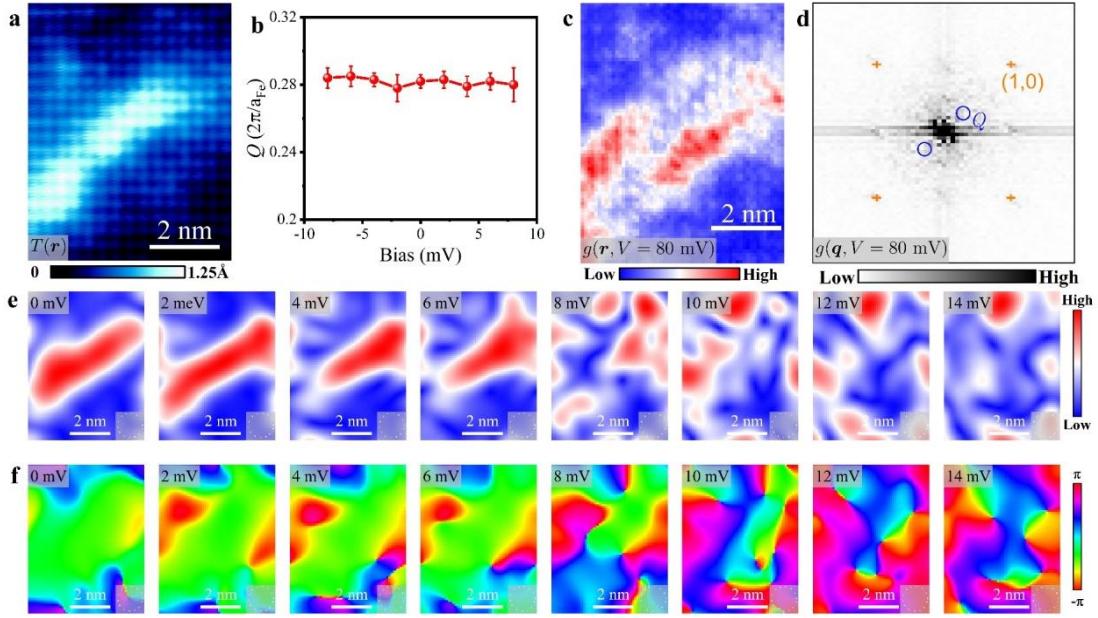

**Extended Data Fig. 3 | Non-dispersive PDW and absence of PDW in high energy d$I$/d$V$ maps. a,** The STM topography of the domain wall in Fig. 4. **b,** The bias dependence of the wavevector of the LDOS modulation at the domain wall. **c, d,** d$I$/d$V$ map (**c**) taken at 80 mV over the same FOV of **a** and corresponding magnitude of the Fourier transform (**d**). There is no spatial modulation in the d$I$/d$V$ map (**c**) and no FFT peak at around $Q$~0.27 $Q_{Fe}$ at the energy much higher than the superconducting gap (**d**). **e, f,** Spatial variation of the amplitude (**e**) and phase (**f**) of the LDOS modulation at energies from 0 to 14 meV. The PDW spatial modulation can be only detected at the energy within the superconducting gap, which favors a primary PDW scenario.

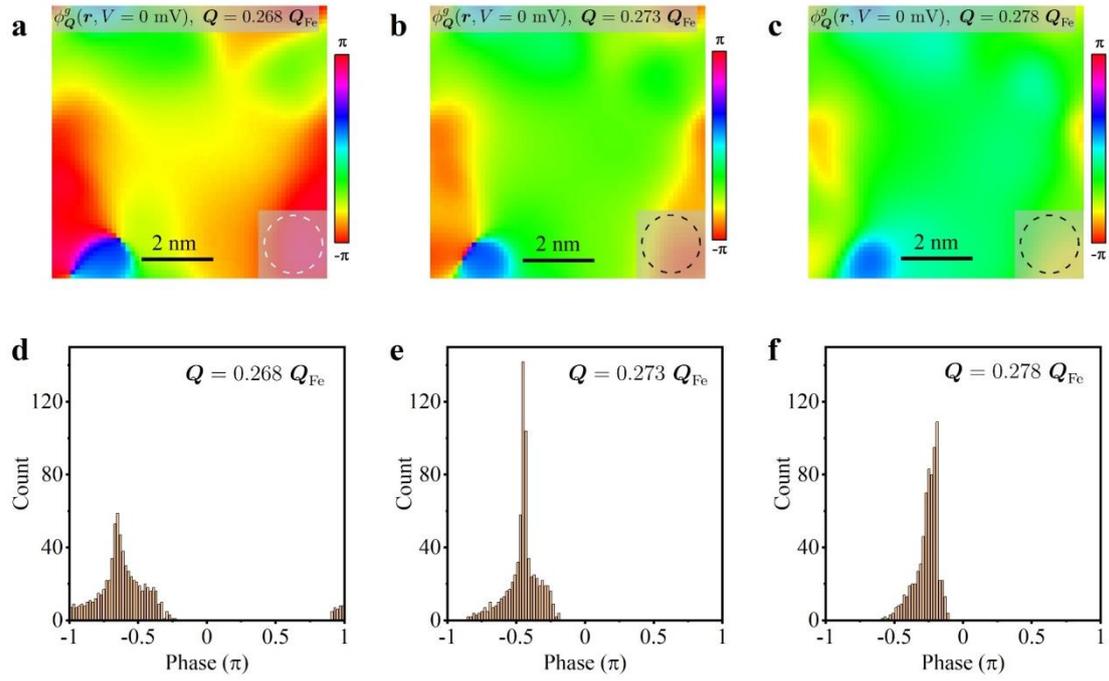

**Extended data Fig. 4 | Accurate determination of modulation wavevector $Q$. a-c,** Spatial distribution of the modulation $g_Q(\mathbf{r})$ phase $\phi^A_{Q_t}(\mathbf{r})$. The averaging length-scales in **a-c** are denoted by dashed circles. **d-f,** Statistical histogram of $\phi^A_{Q_t}(\mathbf{r}_{\text{at the domain wall}})$ at the domain wall for a series of tentative $Q_t$. The determined accurate wavevector $Q$ corresponds to the sharpest peak in $\phi^A_{Q_t}(\mathbf{r}_{\text{at the domain wall}})$ distribution histogram ($\mathbf{r}_{\text{at the domain wall}}$ represents the area of domain wall).

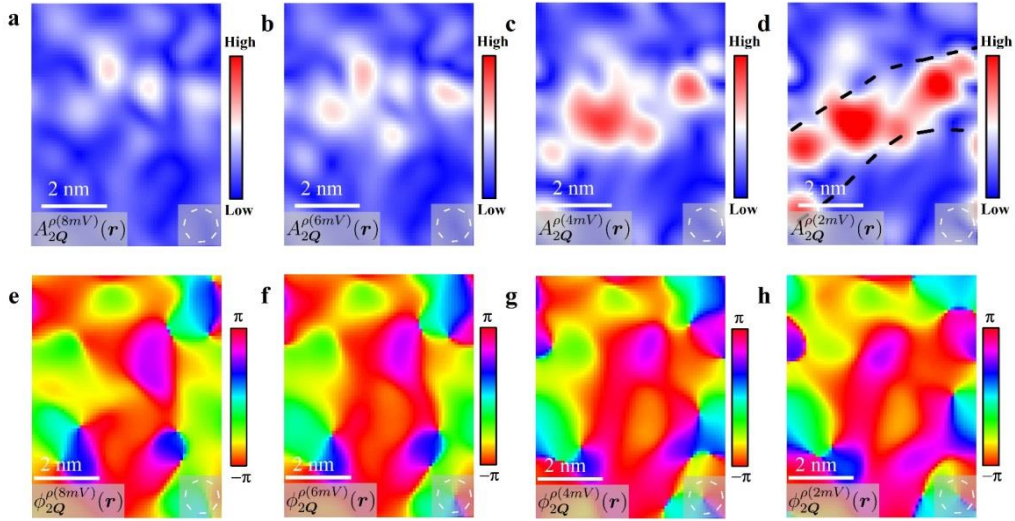

**Extended data Fig. 5 | Spatial variation of the amplitude $A_{2Q}^{\rho(V)}(r)$ (a-d) and the phase $\phi_{2Q}^{\rho(V)}(r)$ (e-g) of the $2Q$ $\rho(r)$ modulation at 8 mV (a, e), 6 mV (b, f), 4 mV (c, g) and 2 mV (d, h)**. The amplitude of the modulation is gradually restricted to the domain wall region and the phase of the modulation becomes more uniform in the domain wall region with decreasing bias voltage, indicating the emergence of the $2Q$ CDW. The domain wall boundary is marked by the dashed line in **d**.

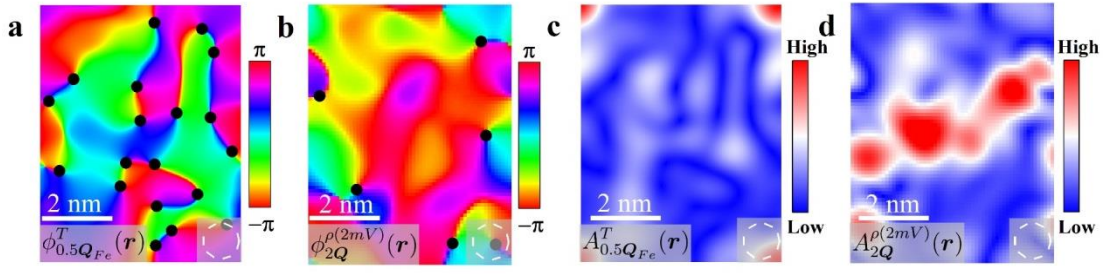

**Extended data Fig. 6 | Comparison between the atomic lattice and 2$Q$ CDW state. a,b,** Spatial variation of the phase $\phi^{T}_{0.5Q_{Fe}}(r)$ of the STM topology $T(r)$ modulation at $0.5Q_{Fe}$ (**a**) and the phase $\phi^{\rho(2mV)}_{2Q}(r)$ of the $\rho(r, V = 2mV)$ (charge density) modulation at $2Q$ (**b**). Topological defects in **a** and **b** are marked by black dots. **c,d,** Spatial variation of the amplitude $A^{T}_{0.5Q_{Fe}}(r)$ of the STM topology $T(r)$ modulation at $0.5Q_{Fe}$ (**c**) and the amplitude $A^{\rho(2mV)}_{2Q}(r)$ of the $\rho(r, V = 2mV)$ modulation at $2Q$ (**d**). The spatial variations of the phase and amplitude show many differences between the atomic lattice and 2$Q$ CDW, indicating that the 2$Q$ CDW is independent of the atomic lattice.

# Supporting Information

# Discovery of a pair density wave state in a monolayer high-$T_c$ iron-based superconductor

Yanzhao Liu[1#], Tianheng Wei[1#], Guanyang He[1], Yi Zhang[2], Ziqiang Wang[3] & Jian Wang[1,4,5,6*]

[1]*International Center for Quantum Materials, School of Physics, Peking University, Beijing 100871, China*
[2]*Department of Physics, Shanghai University, Shanghai 200444, China*
[3]*Department of Physics, Boston College, Chestnut Hill, MA 0246, USA*
[4]*Collaborative Innovation Center of Quantum Matter, Beijing 100871, China*
[5]*CAS Center for Excellence in Topological Quantum Computation, University of Chinese Academy of Sciences, Beijing 100190, China*
[6]*Beijing Academy of Quantum Information Sciences, Beijing 100193, China*

[#]These authors contribute equally.
*Correspondence to: jianwangphysics@pku.edu.cn (J.W.)

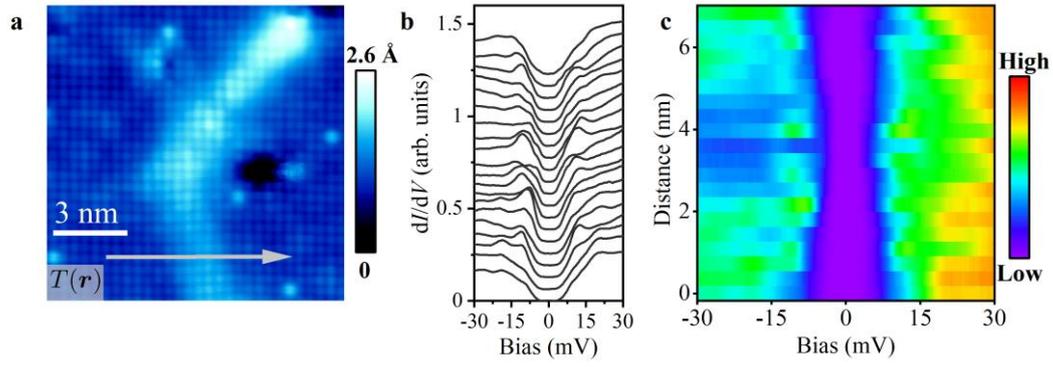

**Fig. S1. Suppression of superconductivity at the domain wall. a,** An STM topography of a domain wall. (12×12 nm$^2$, $V_s$ = 0.1 V, $I_s$ = 0.5 nA) **b,** Tunneling spectra taken along the light grey arrow in **a**. The spectra are vertically offset for clarity. **c,** Intensity plot of **b**.

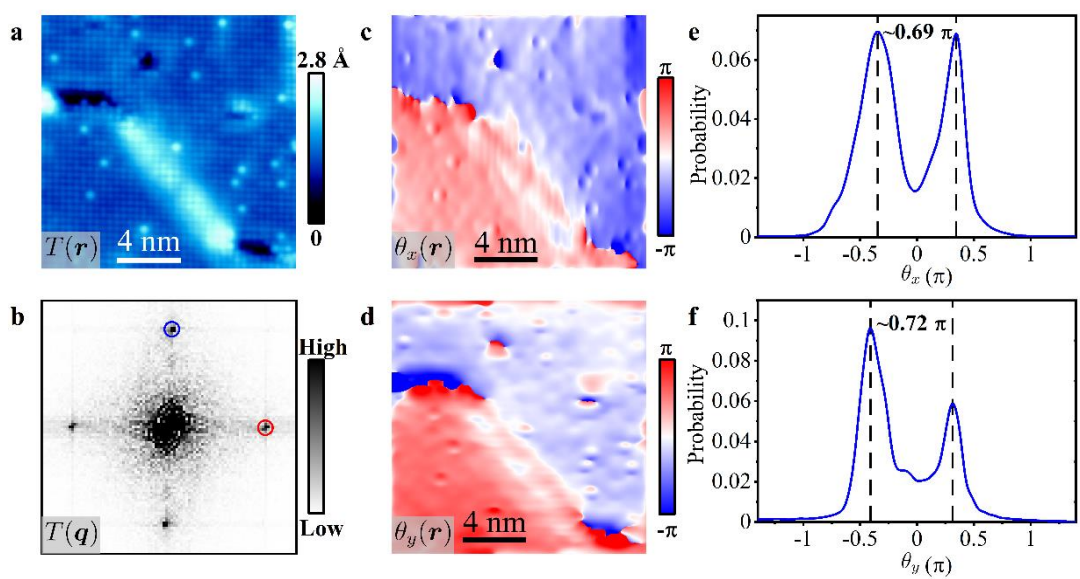

**Fig. S2. Determination of the lattice shift across the domain wall by the 2D lock-in technique.** **a,** The STM image of the domain wall in Extended Data Fig. 2a. **b,** Magnitude of the Fourier transform of **a**. **c, d,** Phase maps of the region in red (**c**) and blue (**d**) circles in **b**, respectively. The phase map is obtained by shifting the region to the center of Fourier transform (**b**) and taking the inverse Fourier transform. **e, f,** Density distribution of the phase in **c** and **d**, respectively.

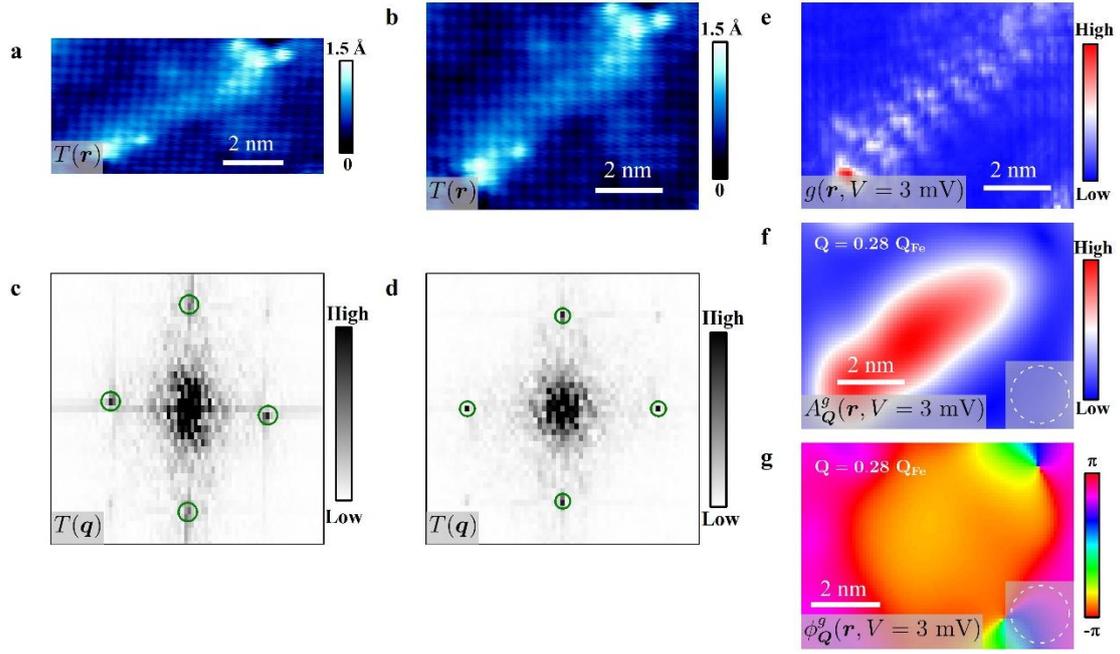

**Fig. S3. Correction on distortions in topography $T(r)$ and differential conductance map $g(r, V)$. a, b,** STM images of a domain wall before (**a**, 9×4.5 nm$^2$) and after (**b**, 8×6.1 nm$^2$) correcting distortions. **c, d,** Magnitude of the Fourier transform of **a** and **b**, respectively. **e-g,** d$I$/d$V$ map (**e**), spatial variation of the amplitude (**f**) and phase (**g**) of the modulation at the domain wall after correcting distortions. The averaging length-scales in **f** and **g** are denoted by dashed circles.

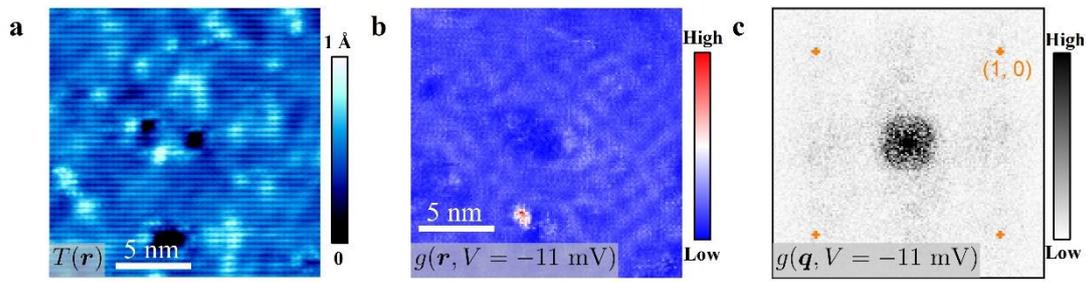

**Fig. S4. d$I$/d$V$ map in the region without the domain wall. a**, STM topography (20×20 nm$^2$, $V_s$ = 0.1 V, $I_s$ = 0.5 nA) of a region without the domain wall. **b, c,** d$I$/d$V$ map $g(r, V = -11$ mV$)$ (**b**) and corresponding magnitude of the Fourier transform $g(q, V = -11$ mV$)$ (**c**) taken in the same region as **a**.

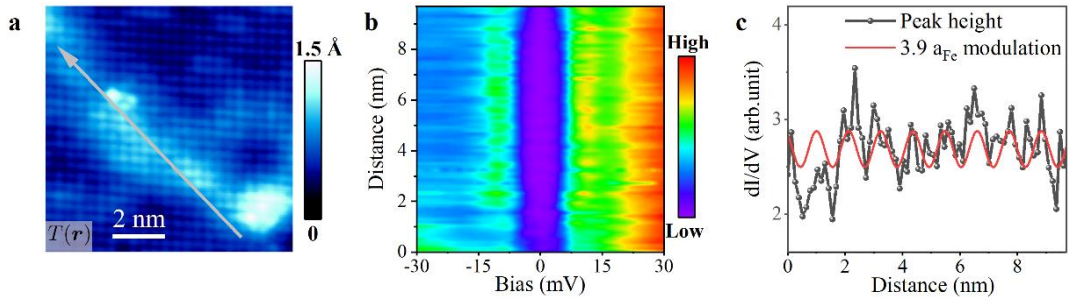

**Fig. S5. Periodic modulations of the coherence peak height at the domain wall. a,** The STM topography (9.5×9.5 nm$^2$, $V_s$ = 0.1 V, $I_s$ = 0.5 nA) of a domain wall. **b,** The line-cut intensity plot along the light grey arrow in **a**. The distance in **b** is defined relative to the beginning of the arrow in **a**. **c,** Measured height of coherence peak in **b**.

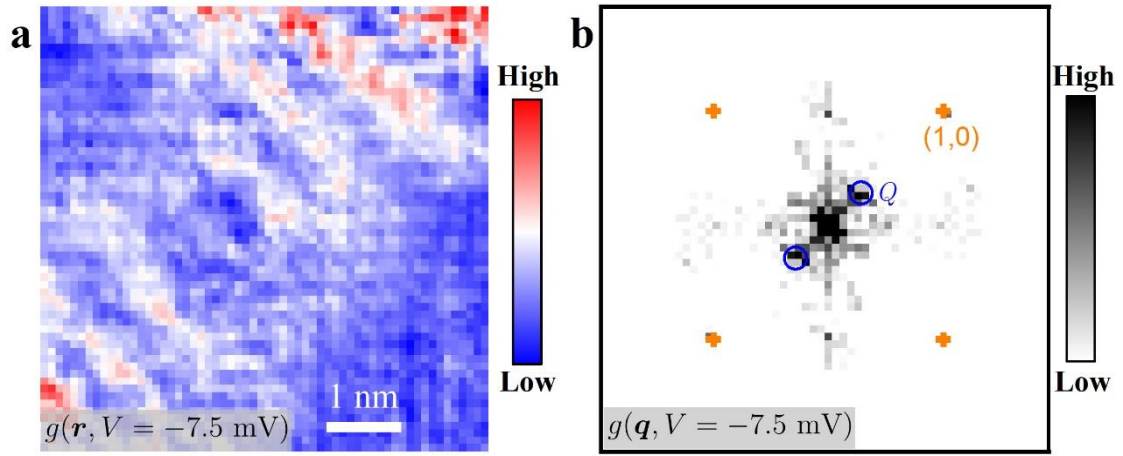

**Fig. S6. Modulation of the LDOS at a bias voltage close to the coherence peak. a,** d$I$/d$V$ map $g(\mathbf{r}, V = -7.5 \text{ mV})$ taken at the same area in Fig. 3 at 4.3 K. $V$ = -7.5 mV is close to the gap edge at the domain wall. **b,** The magnitude of the Fourier transform of **a**. Orange crosses are at $\mathbf{q}$=($\pm\mathbf{Q}_{Fe}$,0),(0,$\pm\mathbf{Q}_{Fe}$). The modulation wavevector $\mathbf{Q}$~0.28 $\mathbf{Q}_{Fe}$ is marked by solid blue circles.

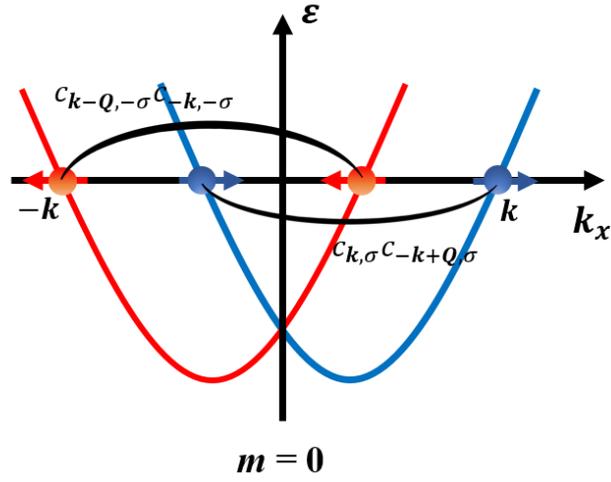

**Fig. S7. Finite-momentum equal spin pairing induced by the *inter*-Fermi surface pairing.** $\langle c_{k-Q,-\sigma} c_{-k,-\sigma} \rangle$ and $\langle c_{k,\sigma} c_{-k+Q,\sigma} \rangle$ are the spin-triplet order parameters with the finite momenta $\pm Q$ in opposite equal-spin pairing channels induced by the SOC. Red and blue curves plot the energy bands ($\varepsilon(k_x)$) with opposite spin orientations which are indicated by red and blue arrows, respectively. Zeeman field ($m$) is zero.

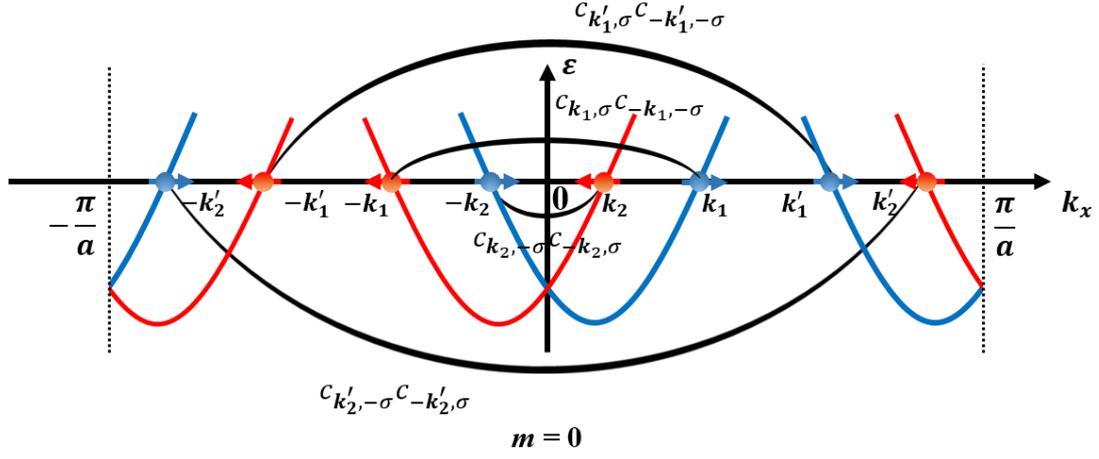

**Fig. S8. Intra-Fermi surface, zero center of mass momentum, opposite-spin pairing.** $\langle c_{k_1,\sigma} c_{-k_1,-\sigma}\rangle$, $\langle c_{k'_1,\sigma} c_{-k'_1,-\sigma}\rangle$, $\langle c_{k_2,-\sigma} c_{-k_2,\sigma}\rangle$ and $\langle c_{k'_2,-\sigma} c_{-k'_2,\sigma}\rangle$ are mean-field opposite spin pairing order parameters, where $\sigma$ is the spin index. Red and blue curves plot the energy bands ($\varepsilon(k_x)$) with opposite spin orientations which are indicated by red and blue arrows, respectively. Zeeman field ($m$) is zero.

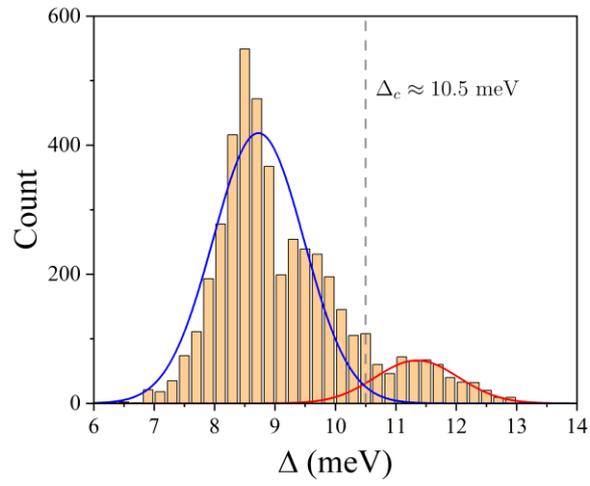

**Fig. S9. The statistical histogram of superconducting gaps extracted from the gap calculation of Fig. 4.** Blue and red curves are the Gaussian fittings.